\documentclass[oneside,reqno,english]{amsart}
\usepackage[T1]{fontenc}
\usepackage[latin9]{inputenc}
\usepackage{geometry}
\usepackage{textcomp}
\usepackage{amsthm}

\makeatletter
\numberwithin{equation}{section}
\numberwithin{figure}{section}

\pdfoutput=1 
\usepackage{etoolbox}
\makeatletter
\patchcmd{\@maketitle}
  {\ifx\@empty\@dedicatory}
  {\ifx\@empty\@date \else {\vskip3ex \centering\footnotesize\@date\par\vskip1ex}\fi
   \ifx\@empty\@dedicatory}
  {}{}
\patchcmd{\@adminfootnotes}
  {\ifx\@empty\@date\else \@footnotetext{\@setdate}\fi}
  {}{}{}
\makeatother

\makeatother

\usepackage{babel}
\begin{document}
\title[On the causal efficacy of qualia]{ On the causal efficacy of qualia: Philosophical zombies are fine-tuned,
and implications for the quantum measurement theory}
\author{Adam Brownstein$^{*}$}
\thanks{$^{*}$Melbourne, Australia. ORCID: https://orcid.org/0009-0001-7814-4384}
\date{February 11, 2025}
\begin{abstract}
\noindent We suggest that qualia have a causally efficacious role
in quantum mechanics; an occurrence which explains how information
about qualia can enter the physical environment. This is compatible
with the unitary time-evolution of the quantum state if qualia are
understood as effecting the beables of the de Broglie-Bohm interpretation
or wavefunction collapse process rather than at the wavefunction level.
We furthermore suggest that not all quantum states are consistent
with qualia. If this is the case, the standard wavefunction collapse
postulates of the Copenhagen interpretation will fail to select only
those states which are consistent with qualia, and the Born-rule must
be modified if wavefunction collapse is to generate the correct dynamical
histories across time. This new model which includes qualia clearly
demonstrates how non-linear and self-referential phenomena can occur,
despite the linear, deterministic time-evolution of the wavefunction.
We reject the notion that physical matter operates independently of
qualia, and find that the main evidence for epiphenomenalism i.e.
the causal closure of the underlying physical time-evolution, has
failed to take into consideration fine-tuning of the microcausal degrees
of freedom. We propose that the philosophical zombie argument is fine-tuned
in the initial conditions, thus making philosophical zombies statistically
unlikely if the fine-tuning is removed. We furthermore suggest the
presence of fine-tuning can be used as a test for consciousness in
the general case. This explains why some classical processes such
as computer circuits may never be conscious; because they lack the
capacity for fine-tuning when the dynamics is reversed.
\end{abstract}

\maketitle

\section{Introduction}

\noindent A very puzzling phenomenon in physics is that our experiences
of qualia, i.e. our internal conscious experiences, appear to have
an effect upon the world. Why this is strange is that the physical
matter has no knowledge of qualia, as it is purely a non-physical
phenomena. As the hard problem of consciousness \cite{key-1} has
shown, the experience of `what it's like to be a brain-state' is distinct
from the collection of neurons which make up the physical brain. This
naturally leads to the question of how information about qualia enters
the physical environment. If this information doesn't come from qualia,
where does it come from? Is it just a convincing illusion? Or perhaps
is there a causal role for qualia?

It is often assumed that consciousness is a physical phenomena, possibly
arising out of self-referential information processing, irreducible
quantum states, or emergent behaviour in the brain. However, these
purely physical explanations cannot address the hard problem, because
the experience of `what it's like to be physical matter' cannot be
explained in terms of physical matter alone. The experience of `what
it's like to be'\footnote{For instance Nagel's ``What is it like to be a bat?' \cite{key-3}}
is an additional property over and above any material or mathematical
explanation. 

It is also often assumed that physics is a closed, causal dynamical
theory. Therefore, any role for consciousness must only be an epiphenominal
one, i.e. an apparent causality. However, this proves to be false
when examined more closely. While quantum mechanics is a causal theory,
it specifies an open-ended probabilistic dynamics, and therefore is
not closed. Therefore, the question of whether the theory is deterministic
is subtle. It is deterministic in some interpretations, such as the
deterministic version of the de Broglie-Bohm interpretation. However,
there is an equivalent stochastic de Broglie-Bohm interpretation,
and the Copenhagen interpretation likewise describes probabilistic
outcomes. These probabilistic or stochastic interpretations indicate
that the dynamics is not closed. 

Taking both facts together, there is an explanatory gap as to how
information about qualia can enter the physical environment; and there
remains a possibility that qualia play a causal role by guiding the
open-ended probabilistic outcome of quantum mechanics. We contend
that epiphenomenalism fails to explain how this information about
qualia arrives into the physical environment, and therefore the explanatory
gap points directly toward a defect in the quantum measurement theory
as defined by the Born-rule. 

To make our philosophical position clear, we fully endorse the existence
of the hard problem of consciousness. For example, it is possible
to describe the quantum mechanics of particles or fields, but the
physics alone cannot explain why `what it's like' to be these particles
or fields is a specific conscious experience rather than simply a
non-conscious state of matter. Therefore, the challenge is to find
a suitable role for conscious experience in the existing scientific
paradigm. 

Taking conscious experiences (qualia) as real, there are only a few
available options. Either qualia are causally efficacious, or they
are an epiphenomenon. An epiphenomenon means that they have no causal
role, but are like a projected image of the underlying reality. If
they are an epiphenomenon, then they are arbitrary, and hypothetically
can be altered in thought experiments. 

These thought experiments which remove or alter qualia pose a significant
challenge to the epiphenomenalist viewpoint. Indeed, everything we
do as conscious human beings is driven by qualia; sight, sound, touch,
internal sensation, cognition, and emotion are all forms of qualia.
If you remove or alter the qualia, it is not clear whether the physical
system can still operate in the same way as it did previously. Our
contention is that epiphenomenalism only works as an apparent explanation
retroactively, looking backward at the previous time-evolution of
a system. To explain the forward time-evolution however, it appears
that epiphenomenalism fails, and qualia must play a causal role. The
distinction between these two cases is the presence or absence of
fine-tuning in the microcausal degrees of freedom. 

In our view, the evidence that information about qualia is transmitted
into the physical world indicates that a causal role for qualia is
a logical necessity. The challenge then is to identify a suitable
mechanism which is compatible with the deterministic time-evolution
of the physical system. This is not as insurmountable as is commonly
believed. For instance, suppose that qualia influenced the wavefunction
collapse process in the Copenhagen interpretation. Would this not
explain both the apparent causal impact of qualia, whilst retaining
the deterministic dynamics of the underlying physics as given by the
Schrödinger equation? This is the type of model which seems to be
necessary to explain how experiences of qualia can impact the real
world i.e. in a top-down causal fashion, whilst simultaneously allowing
room for a reductionist physical description. 

Alternatively, the problem can be understood in terms of the de Broglie-Bohm
interpretation. In this case, the qualia are related to the beables
of the de Broglie-Bohm interpretation, but they have an additional
causal role. This role may either be i) to have a causal influence
upon themselves or ii) to have a causal influence upon the beables.
These two possibilities are compatible with quantum mechanics, because
the dynamics of the wavefunction remains unaffected. 

It is of course indicative that the Born-rule might not be strictly
correct, as it fails to take into account the causal role of qualia.
If the Born-rule were entirely correct, no causal role for qualia
would be required, and epiphenomenalism would adequately explain consciousness.
However as will be discussed, the evidence is suggestive of a causal
role for qualia, thus indicating the insufficiency of the Born-rule. 

Once the causal role for qualia is established, many things become
clear. Firstly, it makes sense of how information about qualia can
enter the physical environment. Secondly, it makes sense of how non-linear
and self-referential effects can occur within quantum mechanics, without
resorting to an explanation in terms of classical mechanics as is
commonly done. Thirdly, it makes sense of whether macroscopic classical
circuits can be conscious. These points will be discussed in detail
in the following sections of this paper, along with other strong pieces
of evidence. 

In regards to the question of consciousness in classical circuits,
we come to some intriguing conclusions. We find that fine-tuning is
a hallmark of consciousness, otherwise the system cannot support causally
efficacious qualia. The lack of fine-tuning in macroscopic classical
circuits may indicate that classical computers as presently envisaged
are not capable of consciousness, even if they have functional input-output
behaviours that simulate consciousness. This goes beyond the Turing
test, because we are able to provide a test for consciousness in terms
of the physical dynamical picture, rather than a system's functional
behaviour. 

The causal role for qualia reopens the question of how organic life
(both conscious and non-conscious) is able to maintain states of high
information content and low entropy. It is likely that consciousness
didn't emerge spontaneously, and biological systems may have evolved
to utilise a basic causal mechanism present in nature which is responsible
for the effect. Therefore, non-sentient matter may also be utilising
the same causal mechanism to support life. 

\section{Models of causation}

\subsection{Model 1: Epiphenomenalism}

One might easily believe that consciousness is all a neurobiological
process. For instance the brain of an individual can perceive colour
by operating a sequence of neuron firings. However, what is concerning
is that the neurons themselves never see the colour, it is held in
the mind's-eye of the individual. While there is a one-to-one correlation
between the pattern of neurons firing and the perceptions of qualia,
there is no explanation of how the information from qualia feeds back
down into the neural network. 

Either there is a direct top-down causal role that the perceptions
of qualia play, or alternatively, we only have the illusion of a causal
role being played. The idea that qualia are an epiphenomenon takes
the position of believing this is only an apparent causality. It is
as if the physical world is projected into our mind's-eye where a
convincing drama plays out.

This is somewhat compatible with the de Broglie-Bohm interpretation,
but not entirely. If qualia are assigned in a one-to-one fashion with
the beables of the de Broglie-Bohm interpretation, then on the surface
there is no issue. The conscious experience will continue to play
the epiphenomenal drama, while the underlying quantum wavefunction
and beables do all the real work of specifying the dynamics. 

However, it doesn't explain how the physical matter is able to react
in a way compatible with our experiences of qualia, despite never
having access to the knowledge derived from those experiences. For
instance, in the de Broglie-Bohm interpretation, the beables (e.g.
particle configuration) have no knowledge of the particular colours,
sounds, emotions or thoughts experienced. How then does information
about these things enter the physical environment? The fact this information
about qualia can be encoded physically, for instance as written or
verbal reports, indicates that the information contained in qualia
feeds back into the physical world.

This information about qualia is not part of the wavefunction. Nor
is it part of the wavefunction collapse postulates of the Copenhagen
interpretation. Therefore, the standard physical description never
contains the information in theory. In practice however, this information
is commonplace. Everywhere around us is the information we have received
from qualia and imparted into the physical world. Either the information
about qualia is just a projection of the information already contained
within the physical matter, or more likely, there is a correlating
mechanism forcing physical matter to conform with our perceptions
of qualia. 

There may be no other way to get this information into the physical
world other than to allow qualia to have a causal role. The question
then is how this can be the case. Viewing the problem using the reductionist
picture in terms of neurons, cells, atoms and subatomic particles;
everything appears causal and deterministic. Quantum field theory
too has shown that nature is deterministic down to the very basic
structure of matter. But does a deterministic physical description
really mean everything has been determined? 

Perhaps a crucial piece of the puzzle has slipped the attention of
the reductionist paradigm; physical phenomena can both have a causal
reductionist explanation as a necessary condition, with an additional
causal influence acting as a sufficient condition. In other words,
the dynamics can be deterministic but open-ended, allowing for an
additional causal influence to close the dynamics. It is only when
determinism is both necessary and sufficient that the dynamics can
be considered deterministic and closed, with no room for additional
causal influences to be injected into the dynamical picture.

Therefore, there is room in the causal reductionist paradigm for a
mechanism that filters information about qualia back down into the
physical time-evolution. This explains both why it appears as if everything
is determined by the initial conditions when looking retroactively
at the time-evolution, and simultaneously where the mysterious information
about qualia arises from when looking at the forward time-evolution. 

\subsection{Model 2: Interactionism}

As stated above, instead of qualia being an epiphenomenon, it is possible
they affect the deterministic dynamics in a causal manner. We suggest
that in the interactionist model, qualia act upon the de Broglie-Bohm
particle configuration to bring about this causal influence. We do
not suggest that qualia affect the wavefunction, which has a closed
deterministic dynamics that is fully specified. 

Top-down causation is indeed compatible with bottom-up reductionist
determinism. The de Broglie-Bohm interpretation itself provides a
specific example of this type of dual causation. It is worth reviewing
the mechanism of the de Broglie-Bohm interpretation further to provide
insight into how a similar strategy can be applied to qualia. 

How the de Broglie-Bohm interpretation works is that the wavefunction
provides the bottom-up reductionist causal picture. But it defines
a multiplicity of possible states i.e. the branches of the wavefunction.
Then, a de Broglie-Bohm particle configuration is able to occupy a
specific branch of the wavefunction. The guidance equation or quantum
potential plays a top-down causal role as it directs the particle
configuration down a specific wavefunction branch. Therefore, the
de Broglie-Bohm interpretation adds a different kind of causality
on top of the wavefunction, yet the whole system remains deterministic.
And now qualia are hypothesized in this paper to add a different kind
of causality on top of the de Broglie-Bohm interpretation, and similarly
we can argue the system remains deterministic. 

The question to ask is ``Does the wavefunction know about qualia?''.
For instance, if a person writes about the blueness of the ocean,
was this part of the wavefunction dynamics? Or did the wavefunction
just set up the preconditions for a possible world in which this occurs.
Indeed, because the wavefunction encodes every possible world, it
sets up the conditions for every possible occurrence to happen, including
this one. 

Clearly, the wavefunction does not know about qualia. The wavefunction
is perfectly described in the reductionist paradigm by the unitary
time-evolution. So then what then does guide this person to contemplate
the blue colour? Perhaps it was the Copenhagen interpretation, which
selects a branch of the wavefunction at random, in which the person
contemplates a particular colour like a card from a deck. Or more
likely perhaps, there were subtle influences arising from qualia,
which guided their branch of the wavefunction in a particular direction
compatible with their experiences of qualia. 

The de Broglie-Bohm interpretation tells a story similar to these
subtle influences about qualia. It says that the probability distribution
can influence the particle configuration down a particular branch
of the wavefunction. However, are these the only influences? Perhaps
other qualia, such as perceptions of image, sound, cognition and emotion
also exert an influence. If there are influences from the qualia,
they can help guide the particle configuration down a different branch,
and in doing so link the experience of qualia into a coherent story.
This is precisely what the de Broglie-Bohm interpretation has achieved
for telling a coherent dynamical story about the probability outcomes,
but a similar coherent story needs to be told about the other remaining
qualia, noting that the Born-rule probabilities can themselves be
considered a type of qualia. 

When framed in this light, we have an explanation of how someone can
perceive colour and write information about it on a page. Observing
the colour caused neurons in their brain to fire in a particular pattern.
This created an experience of qualia in their mind's-eye. The qualia
then caused their de Broglie-Bohm particle configuration to alter
its course down a different branch of the wavefunction; which in turn
caused their neurons to alter their pattern of firing from the default
course; which subsequently caused the person to write about what they
had perceived. 

When viewed only from the perspective of the classical-level description
of neurons firing, it all seems deterministic with no room for a causal
explanation. Therefore, one might be mislead to believe the reductive
physical explanation has told the whole story. What is missing is
the fact that the top-down causal influence is acting like a switching
mechanism, rather than as an overt influence upon the motion. The
switching mechanism enables the particle configuration to choose between
the different branches of the wavefunction, much like a train which
is able to switch between different branching tracks. Each branch
of the wavefunction describes a deterministic time-evolution, therefore
looking backward at a particular trajectory it appears deterministic.
However, additional causal influences are able to affect which branch
is chosen when it comes to the bifurcation points. 

\subsubsection{Quantum interpretations }

Alternatively to using the de Broglie-Bohm interpretation, one could
equivalently say that the qualia affect the wavefunction collapse
process of the Copenhagen interpretation. We recognize that every
statement made within the de Broglie-Bohm interpretation has an equivalent
statement in the Copenhagen interpretation, and within most quantum
interpretations compatible with the Born-rule probabilities. Therefore,
the arguments proposed within this paper are relevant to other quantum
interpretations. We use the de Broglie-Bohm interpretation as it provides
a particularly clear formulation of the problem. However, it is possible
to translate these statements to other interpretations as required. 

\subsection{Model 3: Parallelism }

An alternative model is that qualia only have a causal impact upon
themselves. In this approach, the qualia may receive information from
the beables of the de Broglie-Bohm interpretation, but have a separate
dynamics of their own. This is similar to how the beables have a separate
motion, but are related to the wavefunction by a novel mechanism.
The information received would act as correlating mechanism which
forces to qualia and beables to match, just as the guidance equation
is a correlating mechanism forcing the beables and wavefunction to
match. 

It is interesting to note that if one examines the logic of the de
Broglie-Bohm interpretation closely, it resembles the model of psychophysical
parallelism as proposed by Leibniz. The de Broglie guidance equations
ensure that the beables match the wavefunction, thus creating the
`pre-established harmony'. What was not recognised in the model was
that a wavefunction could set up multiple possible branches, so it
is not necessary to correlate the dynamics between the material and
non-material worlds in advance. In other words, the de Broglie-Bohm
interpretation has the `harmony' component (i.e. correlating mechanism),
but not the `pre-established' component of the statement. Setting
up many branches of the wavefunction serves as a replacement for correlating
the two time-evolutions in advance. 

\subsubsection{Many Born-rule worlds}

The general model of parallelism can also be extended to qualia. Instead
of a single de Broglie-Bohm particle configuration or single set of
beables, it is possible there may be an ensemble of them. This would
create a multiplicity of possible Born-rule worlds, similar to how
the wavefunction sets up multiple branches (which contain both Born-rule
and non Born-rule worlds). 

In this scenario, instead of directly affecting the motion of the
beables, the qualia then can have an independent motion. The qualia
will be able to select which set of beables they choose to replicate,
and there is possibility for subtle causal influences to guide their
path. 

This is an intriguing proposal, because it views the de Broglie-Bohm
interpretation on the same level as a Many-Worlds interpretation for
the Born-rule worlds defined by the beables. In fact, it may be philosophically
preferable, because of the conceptual barrier between the material
and non-material worlds. 

To explain the conceptual barrier further, one might expect that all
causal influences which affect the material world can be described
mathematically, since the material world itself has a mathematical
description. And furthermore, one might expect that all things which
can be described mathematically can be represented as part of the
material world. Therefore, it is difficult to understand how a causal
influence can cross from the non-material world of qualia to the material
world, without itself being represented mathematically and becoming
part of the material world. 

If the de Broglie-Bohm particle configuration is material, then it
is difficult to understand how a causal influence from qualia can
cross the divide. However, by having an ensemble of de Broglie-Bohm
beables, qualia only need to influence their own dynamics by selecting
a specific set of beables to replicate; and therefore the causal influence
does not need to cross the divide. However, we do note that a) perhaps
our understanding of this conceptual divide is limited, and causal
influences possibly can cross b) perhaps beables are on the non-material
side of the divide, which would obviate the issue or c) perhaps the
causal influence is a non-physical effect, similar to the anthropic
principle, which is an effect of post-selection on the state. 

\section{Arguments against epiphenomenalism}

\noindent There are two main strategies for explaining conscious
experience. Firstly, it may play a causal role. Secondly, it may be
an epiphenomenon. In the previous section, we primarily focused on
how consciousness can play a causal role. In this section, we will
focus on why it cannot be an epiphenomenon.

\subsection{Arbitrariness}

A first class of argument against epiphenomenalism is that if consciousness
has no link to the underlying physical dynamics, then it is arbitrary.
Why not then change the colour red to green, or make even more dramatic
changes? No changes to the perceptions of qualia will affect the time-evolution
of the physical matter. 

\subsubsection{Rebuttal }

To rebut the arbitrariness argument, it might be said that the qualia
are uniquely determined by the physical matter, like a film projection
which is uniquely determined by the roll of film which is played.
This is a valid argument to make, and therefore it is not useful to
consider thought experiments where the qualia are partially altered
e.g. inverting red to green. These changes may simply give nonsensical
accounts of qualia, which violate the uniqueness assumption of the
mapping from the physical states to qualia. 

This is to say it is possible to take an ``all or nothing'' approach
to qualia i.e. states either have the full set of qualia or none.
By taking an ``all or nothing'' approach, the only remaining arbitrariness
arguments against epiphenomenalism come from philosophical zombie
thought experiments. In these thought experiments, the hypothetical
alteration made to the qualia is to make it completely absent. We
argue that the philosophical zombie thought experiments display information
about qualia entering the physical picture from an unknown source,
which is identified to be the fine-tuned microcausal degrees of freedom. 

\subsection{Corollary: Brain-region replacement }

There is one notable exception to the ``all or nothing'' approach.
While it is not possible to arbitrarily change the qualia assuming
the one-to-one mapping, it is possible to change the physical picture
of the brain without affecting the functional input-output relations
of its operation, and this change in the physical picture may in turn
change the qualia. This leads to thought experiments where individual
brain regions are replaced by non-conscious circuits, for instance
replacing a particular brain region with a computer chip.

Assuming the classical computer circuit is not capable of consciousness,
making this change might produce a partial removal of some qualia,
while preserving the overall physical operation of the brain. If these
thought experiments produce logical contradictions, they provide evidence
against epiphenomenalism. The fate of epiphenomenalism would be forced
to hang in the balance of whether classical computer circuits are
conscious or non-conscious, or at least whether they can be integrated
along with biological neurons in the formation of qualia. 

\subsubsection{Replacing the color-perception region}

Consider a thought experiment where the brain region responsible for
colour perception is replaced by a classical computer circuit, thus
potentially removing the qualia of colour perception. Suppose this
person is shown a green and a red object, and asked to choose which
one is green. They may have a sense of the green object, because the
functional information processing capabilities of their brain are
unchanged. However, because colour perception is no longer part of
their qualia, they will not be able to directly observe the colours
in their mind's eye. Therefore, the sense of the green colour will
only be subconscious. This seems to be contradictory, because a basic
property of consciousness is that it is known when our decisions are
based upon a conscious awareness of our qualia, and when they are
based upon gut feeling or subconscious reasons outside of our awareness. 

It is as if our conscious experience sets up a barrier between self
i.e. that which is within conscious awareness, and non-self i.e. that
which is outside of conscious awareness. For instance, when a decision
is made based upon gut feeling, we may have a sense that it is the
right decision, but we do not know precisely why we have that sense.
If instead a decision is made based upon our qualia, we know precisely
why we have made the decision, and have an understanding of the causal
factors and qualia involved. 

\subsubsection{Implications }

If one assumes that qualia play no causal role, it is difficult to
explain why the two scenarios (of having a completely biological brain
or a brain partially replaced by a computer chip) are different. A
brain region which has been replaced with the artificial circuit should
theoretically behave exactly the same as an organic brain in terms
of functional operation. Yet it is evident that the removal of the
qualia of colour from the individual by making this replacement will
result in different behaviours. Their perception of causal completeness
of their qualia will be altered. Therefore, we conclude that these
thought experiments produce a contradiction. It is evidence that qualia
do indeed play a causal role.

\subsection{Information causality}

The second major class of argument against epiphenomenalism is that
if epiphenomenalism is true, information about qualia enters the physical
world with no known causal mechanism. Our perceptions of qualia alter
our actions, which causes the information to be encoded from our experiences
into the arrangement of physical matter. If one denies this causal
role, then how does this information enter the physical environment?
There are several peculiar aspects to this: 
\begin{enumerate}
\item Physical reality always seems to be biased toward encoding the truth
value of statements about qualia. Yet the physical matter has never
observed the qualia directly. For instance, a person could be asked
a binary question about their qualia such as ``Is your experience
of qualia continuous?'', to which they answer either yes or no. When
answering this question, somehow the word yes rather than no always
becomes part of the physical world. Yet there is no explanation for
how this information enters the physical world, or what this information
pertains to, if epiphenomenalism is true. 
\item This information about qualia has to come from somewhere. However,
if one looks at the quantum description of the time-evolution, it
is not present there. The initial quantum state does not contain it.
The time-evolution operators do not contain it. Therefore, the time-evolved
wavefunction does not contain it. The information is not a part of
the paradigm of quantum mechanics at all. 
\item We suggest that when this information is not properly accounted for
(e.g. by abstracting qualia away from the physical dynamical picture)
it gets shifted into the fine-tuned microcausal degrees of freedom,
and will be present in the initial state of the de Broglie-Bohm particle
configuration for example. The presence of fine-tuning is theoretically
very undesirable, as it means the initial state a) contains future
information and b) lacks robustness to specification. 
\end{enumerate}

\subsubsection{Rebuttal}

A rebuttal can potentially be made that if qualia are an epiphenomenon
this may only be an apparent causality. From this viewpoint, the answer
to the question of where the information came from is that there is
a one-to-one correlation between qualia and the physical brain states.
Therefore, the information is already present in the physical world,
just hidden in a different form. This argument falls apart however,
because it is not clear how the physical environment ever contained
the information in the first place. It is just another way to suggest
that the microcausal degrees of freedom of the initial state contain
hidden information, and therefore it is indicative of fine-tuning. 

There is a second way to explain the situation using epiphenomenalism.
Perhaps the information is never part of the physical environment,
only apparently so. For instance, perhaps the information is always
contained within our mind's perception and interpretation of the physical
environment, and is never actually transmitted to the physical environment.
For instance, if the word yes is written rather than no, this only
has a semantic meaning in our mind; so it is more a matter of how
the conscious mind assigns meaning to physical states of the environment. 

This is a somewhat convincing account, yet it is not correct to state
that the physical matter never contains the information. Certainly,
the physical world itself is not able to interpret the information
carried by the arrangement of physical matter, but the physical matter
acts as the storage medium and carrier of the information. This is
similar to how words written on a page store information, yet the
words are only meaningful when read by a person. Requiring a person
to read the words doesn't detract from the fact that the information
is contained in the arrangement of words on the page. 

Furthermore, the information about qualia imparted into the physical
world can i) be shared and transmitted to other individuals ii) be
copied and replicated and iii) has meaning in relation to other information
derived from qualia. This is all suggestive that the information actually
is imparted into the physical environment. 

\subsection{Corollary: Spontaneous emergence of statements about qualia}

\subsubsection{``Zombie island'' thought experiment }

Imagine a tribe located on a remote island, where each member of this
tribe are all philosophical zombies from birth. Nobody on the island
has ever experienced qualia. This tribe develops language, culture,
writing and philosophical reasoning independently from the rest of
the world. Because the physical description is identical to the case
where they do have qualia in the epiphenomenalist paradigm, they supposedly
start describing experiences of qualia. They communicate properties
about their qualia; for instance that it is continuous in nature,
that it has a single locus of identity instead of several, that they
can see colours and feel sensations. Yet they have never actually
experienced this qualia.

Where did all of this information about qualia arise from? It didn't
come from within them, because they were born without qualia. It didn't
come from past culture, because nobody in their prior culture ever
experienced it. And it didn't come from the environment either, for
their environment only contained non-sentient matter.

Apparently, the information must have come from the pattern of neurons
firing in their brain. However, it would be extremely unlikely for
a closed system such as this to develop descriptions for what qualia
is like. This indicates that something is wrong in the physical description
of this system. Knowledge about qualia cannot spontaneously arise
in the physical world without some causal reason or origin of the
information. 

\subsubsection{Free and counterfactual philosophical zombie arguments}

In actuality, there is no evidence that this is how the dynamical
time-evolution will play out. If the physical system undergoes free
time-evolution in isolation from conscious experience, it is not clear
that the philosophical zombies would be able to function similar to
real human beings. 

The typical story told about philosophical zombies is based upon the
assumption of having a reference system, and removing the non-physical
aspects of qualia from the picture. However, this procedure is questionable,
as the end result is a set of physical particle trajectories that
have co-evolved with qualia, thus the trajectories have been post-selected
on the existence of qualia. To give an accurate account of the physical
situation in the forward-directed sense of the time-evolution, it
is necessary to distinguish between free time-evolution and counterfactual
time-evolution that has been post-selected. 
\begin{enumerate}
\item Counterfactual (post-selected) time-evolution: This is the traditional
philosophical zombie argument. Take a dynamical system (reference
system) and split the material description from the non-material component.
Only look at the material description in isolation. 
\begin{enumerate}
\item It is possible to do this, because the de Broglie-Bohm beables ensure
that a deterministic material description of the time-evolution exists
when looking at the counterfactual, post-selected state. Therefore,
philosophical zombies are possible. Because philosophical zombies
are possible, it demonstrates that consciousness is an additional
fact which needs to supposed in addition to the material description.
Thus the hard problem of consciousness is established. 
\item However, the deterministic state has potentially been influenced by
qualia, as it has co-evolved with the presence of qualia. The initial
state (e.g. the initial positions of de Broglie-Bohm particles) has
been influenced by specifying the final state (post-selection) owing
to the deterministic paradigm which provides a one-to-one mapping
between initial and final states. Choosing the final state also determines
the initial state due to the deterministic time-evolution, which results
in fine-tuning.  
\item To understand this further, imagine the dynamics as a branching process.
Looking backward, all of the emanating branches converge to a root
branch, which causes the dynamics to appear causally closed and fully
determined. Looking forward however, the diverging branches make the
dynamics open-ended and requiring further specification. It is easy
to miss the fact that there was a causal influence guiding the choice
of branches along the way if one only looks at the backward time-evolution.
All of the information about the causal influences from qualia will
become shunted into the microcausal degrees of freedom e.g. fine-tuning
of the initial state of de Broglie-Bohm particles, if the final state
is post-selected and qualia are ignored. 
\end{enumerate}
\item Free time-evolution: This is the ``zombie island'' thought experiment.
The physical description is described separately from the non-physical
phenomena from the outset. There is no reference system, and the dynamical
time-evolution takes place completely de novo. Furthermore, ideally
no conscious entities are present in the initial state. Therefore,
the system has never encounters qualia at any stage. 
\begin{enumerate}
\item This system will fail to develop spontaneous emergence of information
about qualia. This is because the microcausal degrees of freedom have
not been fine-tuned by a counterfactual, post-selection procedure.
\end{enumerate}
\end{enumerate}

\subsubsection{Analogy to the anthropic principle}

A useful analogy can be made to the anthropic principle. When one
looks backward (counterfactually) at the previous time-evolution of
the cosmos and life on Earth, it can be seen that upon each step of
the process, there was a deterministic causal explanation for it.
However, the events which unfolded were extremely unlikely to occur
statistically. The anthropic principle has been a guiding force which
has enabled the statistical likelihoods to be overcome. The anthropic
principle plays a quasi-causal role, forcing the statistics to generate
a particular outcome, due to post-selection of the final state. 

A similar situation is occurring in the philosophical zombie thought
experiments. Philosophical zombies that can make statements about
qualia are extremely unlikely to occur naturally in the free-time
evolution. However, the statistical likelihoods can be overcome due
to the influence of qualia along the way. Ignoring the guiding influence
of qualia and unknowingly imposing post-selection upon the final state
results in seemingly paradoxical phenomenon such as the apparent spontaneous
emergence of statements about consciousness. 

\subsubsection{Generality }

Just as the hard problem of consciousness says that there can be no
purely physical explanation of consciousness experience, we believe
that the spontaneous emergence of statements about consciousness (e.g.
in the ``zombie island'' thought experiments) indicates that there
can be no purely epiphenominal explanation. Therefore: 
\begin{enumerate}
\item Physicalism is false (as demonstrated by the hard problem of consciousness).
\item Epiphenomenalism is false (as demonstrated by spontaneous emergence
of statements of consciousness).
\item Therefore, a causal role for qualia is true. 
\end{enumerate}

\subsection{Self-referential information}

A third class of argument against epiphenomenalism is that conscious
beings observe and act upon self-referential information. Yet the
underlying dynamics provided by the quantum formalism is deterministic
and linear, so has no capacity for self-referential behaviour. 

\subsubsection{Looking in the mirror}

If the wavefunction undergoes linear time evolution, how can a person
see themselves in the mirror and then act upon their observation?
What it implies is that there is a feedback mechanism between qualia
and physical matter. How it works is roughly as follows: 
\begin{enumerate}
\item Receiving step: Photons are emitted from the surface of the human
body and reflected in the mirror. The eye then receives the photons,
which causes a signal to be transmitted to the brain. This causes
the wavefunction to branch into many different brain-states. The de
Broglie-Bohm particle configuration occupies a specific branch of
the wavefunction corresponding to a particular brain-state (i.e. pattern
of neurons firing). The specific configuration of de Broglie-Bohm
particles also causes a specific qualia of sight to occur in the mind's-eye
of the individual, due to a mapping between the de Broglie-Bohm particle
configuration and the qualia experienced. 
\item Emitting step: The qualia will then cause the de Broglie-Bohm particle
configuration to change branch. The fact the de Broglie-Bohm particle
configuration has changed branch will cause the neurons to fire in
a different pattern than would have otherwise occurred, causing motion
of the individual's body that is reactive to the visual stimulus. 
\end{enumerate}
In this new explanation, there is a recursive feedback loop between
the physical environment and the qualia, enabled by its impact upon
the particle configuration. This enables self-referential information
to impact the dynamics of physical matter.

\subsubsection{Copenhagen interpretation}

The above picture is in stark contrast to the story told by the Copenhagen
interpretation. The explanation provided by the Copenhagen interpretation
is approximately as follows: ``The wavefunction continually collapses
into a classical state. Although the quantum time-evolution is entirely
linear, because there is a classical explanation of the brain process
in terms of neurons firing, it is explainable as a self-referential
loop in terms of this classical-level description. Quantum mechanics
is a linear theory, but wavefunction collapse makes it non-linear.''.
This story is problematic because a) it evokes classical mechanics
to explain the situation. Essentially, it is an admission that quantum
mechanics fails to explain the phenomena, and b) it assumes the classical-level
description is deterministic (which is correct) and not fine-tuned
(which is incorrect). 

To see what is wrong with the explanation, the Schrödinger time-evolution
can be continued for longer in a no-collapse interpretation of the
wavefunction. This will not introduce self-referential, non-linear
information into the physical description because the Schrödinger
time-evolution is linear. After a long period of time evolution, the
quantum state will contain decoherent branches representing the variety
of possible classical worlds. Although the states of the world in
which this non-linear behavior occurs are present in the wavefunction,
they will have small wavefunction amplitudes and are unlikely to be
selected by the Born-rule.

\subsubsection{Number of cycles}

Note that the greater number self-referential cycles that occur, the
greater the deviation between the probability for the real state observed
and the probability amplitudes described by the quantum formalism.
This is why self-referential loops pose such a problem for quantum
mechanics. If there is one cycle of emission of photons and reflection
from the mirror, the wavefunction amplitudes may deviate a small amount.
However, if one stands in front of the mirror for some time, a constant
stream of photons are emitted and reflect from the mirror. The self-referential
loop will have many cycles, which will cause a large deviation of
the real probabilities from the quantum amplitudes. 

Typically, in such situations it is assumed that the wavefunction
continually collapses to the classical state, so that the self-referential
behavior can be accommodated in the classical framework. Yet, continually
evoking the wavefunction collapse procedure hides the truth that the
quantum amplitudes accumulate small deviances from the real probabilities
over time under the linear time-evolution. 

Each time wavefunction collapse is used, the probability amplitude
information is discarded. Therefore, these small differences can become
hidden by invoking continual wavefunction collapse. The effect might
be considered as related to the quantum Zeno effect, which observes
that continuous measurement of a quantum system alters it to behave
classically. In general, we postulate that the causal role of qualia
may be to have a quantum Zeno-like effect, or an effect similar to
intrinsic decoherence. 

\subsection{Information processing of the brain}

A fourth class of argument comes from examining the information processing
capabilities of qualia. There are proposals which state the brain
is able to perform computational tasks that exceed the capabilities
classical mechanics. It has been suggested this might be due to quantum
structures in the brain that survive decoherence. However, alternatively
it is possible that the information processing power of the brain's
software (for example qualia solving the binding problem of generating
a unified experience) might exceed the information processing power
of the brain's hardware (a network of physical neurons which do not
solve the binding problem).

The information processing required to construct a coherent three-dimensional
image for example, might not be entirely due to the neural network.
It might be a result of the information binding capabilities of qualia.
The physical description of the brain alone i.e. the neural network;
even if quantum mechanical in a limited capacity; might not contain
the real power. Not much is known about qualia, but given the fact
quantum states are capable of high-dimensional, parallel information
processing, it is not out of the question that qualia also are able
to access high-dimensional, parallel or otherwise novel information
processing capabilities. Furthermore, the qualia itself may be a quantum
mechanical effect, only on the non-material side of dualism.

The proposal that qualia are responsible for the information processing
power of the brain makes more sense than claiming the advanced information
processing capabilities are due to quantum structures that survive
decoherence. In particular, the qualia remain relevant even if decoherence
is dominant in the brain. 

\subsection{Corollary: Novel qualia }

Similar to the argument for information processing power of qualia
(a unique capability), it can be argued that the qualia have unique
properties not predicted by the underlying physical brain-state. If
this is true, when this unique information is conveyed into the real
world, the only possible source of the information is from qualia,
thus implying a causal role. Therefore, it should be possible to examine
the properties of qualia and search for unique properties which cannot
be predicted by the underlying physical matter. If such properties
exist, it provides evidence that qualia play a causal role.

\subsubsection{The richness of experience }

The richness of the experience of qualia doesn't seem to be captured
by the physical representations. For instance, ask a philosophical
zombie what their favorite colour is. How can they formulate an answer
without seeing the colours? The standard answer is that the perception
of colour plays no causal role, because whatever is triggering a response
to the question is already explained by the physical description of
the neurons firing in the brain. However, as we have explained, the
simultaneous physical description is already impacted by the actual
presence of qualia information, and it is impossible in principle
to separate the two without fine-tuning the microcausal degrees of
freedom. Therefore this argument is invalid and provides no evidence
for epiphenomenalism. We are forced to confront the fact that the
richness of experience of qualia might not be able to be reductively
explained by physical brain states. 

\subsubsection{Imagined colours}

Another interesting example of novel qualia can be found in the space
of colour perceptions. It is easy to understand how perceptions of
colour could correspond in a one-to-one fashion with the spectral
frequencies of light. However, it is not clear why the brain invents
new colours out of combinations of different frequencies. For instance,
the colour purple is not found in the visible spectrum of light. Purple
is actually produced in the mind of the individual when shown a combination
of red and blue. There are several perplexing aspects of this: 
\begin{enumerate}
\item It is not obvious why qualia should a priori identify blue with red
light to form purple. The brain has connected two opposite ends of
the visible spectrum, where red corresponds to low-frequency light
and blue high-frequency light, and has manufactured purple to close
the loop. Connecting the colour spectrum to form a closed loop in
this way might not be a unique choice. 
\item Furthermore, the perception of the colour purple has nothing to do
with the learning process of a neural network. In our natural environment,
there are many green, blue and brown colours; but vivid purples are
not often encountered.

It is easy to conceive that a brain in the natural environment has
no prior knowledge of purple. However upon seeing purple for the first
time, the brain instantly sees it as purple and this presumably happens
instantaneously in a similar way for most individuals. How does the
neural network of the brain instantly know how to map blue with red
to form purple, across all individuals in a highly similar fashion?
It seems to be a fact only obtainable outside the model of the brain
as a physical learning machine, and rather comes directly from our
qualia. 
\end{enumerate}

\subsubsection{Mapping the space of qualia: Qualia and geometrical/topological information}

Indeed, the whole spectrum of colour perception can be mapped out,
where it forms an interesting geometrical structure. This is perplexing
because what would this mapping represent if not a direct mapping
of our qualia? Generating this mapping without actually experiencing
the qualia does not seem possible. Furthermore, it is curious that
this mapping, which forms a topological space out of the colour combinations,
occurs in a similar same way for most individuals; a fact which can
be studied experimentally by asking participants about their subjective
experiences of colour.

We suspect that other qualia, for instance sound, smell and taste
have similar properties regarding their topological mappings. If these
mappings were due to the way the neural network became wired through
a gradual learning process, it would be unlikely that each individual
brain would independently represent the information in the same way.
One would expect that different neural networks would learn to map
the information in different ways. 

\subsubsection{Artificial qualia: Paradoxical orange }

Another phenomena involving colour are the so called impossible colours.
Impossible colours are those which are not part of the ordinary colour
experience. They can be generated by altering the photoreceptors from
their natural state. Take for example, the colour of `paradoxical
orange'. Paradoxical orange can be generated by staring at a cyan
image until the photoreceptors for this wavelength become saturated.
Then, removing the original cyan image, an orange afterimage appears.
This orange afterimage is perceived to be more dense than any natural
orange to which it is compared. 

You wouldn't suspect a priori that paradoxical orange would occur
in this arrangement. There is no logical reason that the phenomenon
is experienced in the way that it is. The experience seems to come
directly from the way the qualia colour our brain-states. Yet individuals
are able to discus the phenomenon, and their perceptions of it alter
their actions in the real physical world, for instance by making descriptions
of it. 

\subsection{Evolutionary argument}

A fifth class of argument can be made with respect to possible evolutionary
evidence. Biological organisms may have evolved to enhance and utilize
qualia, not merely process information using physical brain computations.
The organizational structure of the brain may actually be below optimal
as an information processing network, but optimized to function more
as an antennae for receiving perceptions of qualia. 

One would imagine that whether the brain fundamentally performs computations
or instead receives computational results from an external source
(i.e. from the non-material qualia) would affect its structure and
function. This is a novel argument, because it can be studied in actuality
by examining the brain and thinking carefully about its organization
and operation. Furthermore, we have analogue models in the form of
artificial neural networks, whereby it might become apparent that
the brain that acts more as a receiving device rather than a computational
device in contrast to artificial neural networks. 

The impact of the brain's information processing style may be examined
in the evolutionary history or across different species. The presence
of qualia may have imposed a strong selection pressure toward enhancing
the brain's capabilities for receiving qualia information at the expense
of alternative physical-based mechanisms of information processing.
This connects the fields of physics, philosophy and evolutionary biology
in a novel way. 

\subsection{Entropy argument }

A sixth class of argument against epiphenomenalism is that type of
information we impart on the physical world due to qualia appears
to be of a different kind than that contained in the motion of physical
matter. Consciousness is able to achieve a global picture of a situation,
and understand the arrangement of physical matter and objects in the
sense of relational information. Deterministic reductionist dynamics
is not relational in this manner.

Using this relational information, it is possible to exert active
control to remove entropy from subsystems and shift it into the external
environment. Conscious perceptions essentially lets us play the role
of Maxwell's demon, shifting entropy from one side of a divide to
another. It is not clear where this capacity for manipulating relational
information and controlling the flow of entropy arises from, if everything
is described by the linear time evolution of the wavefunction. 

It is often assumed this is just due to the non-linear classical-level
description of information processing. However, it is peculiar that
classical mechanics is the only way we have to understand the phenomenon
when the world is fundamentally quantum mechanical. Resorting to a
classical-level explanation seems to indicate that we do not truly
understand the phenomenon as much as commonly believed; and it may
hide the uncomfortable truth that in the classical-level description,
the microcausal degrees of freedom are fine-tuned, and contain hidden
information which has been imposed by post-selection of the final
state. 

\section{Arguments for epiphenomenalism (and why they fail) }

\subsection{The one-to-one correlation argument}

The strongest case for epiphenomenalism comes from suggesting there
is a one-to-one correlation between the material states of physical
matter and non-material states of consciousness. This means that any
causal role is possibly just an apparent causation. For instance,
we have described this interpretation as regarding the physical states
like a roll of film, and the conscious states as a projection of the
film which tells a story of the underlying matter, and is in a one-to-one
relationship with it without affecting the underlying physical dynamical
picture. 

\subsubsection{Counterargument 1: Begging the question}

As discussed in the previous section, it is not possible to separate
the physical time evolution from the qualia without begging the question
(i.e. assuming that which is to be proven) and assuming from the outset
that the qualia are irrelevant. The situation is that of the distinction
between the counterfactual, post-selected time-evolution as compared
to the free-time evolution. If one proceeds and performs this separation
of qualia from the physical picture ignoring this fact, it forces
the microcausal degrees of freedom to become fine-tuned, thus merely
hiding the potential influences from qualia. 

Why this is so problematic, is that even if qualia had a causal role,
if you look back at the historical time-evolution of the system, you
will see a one-to-one relationship that makes it appear as if it had
no causal role. Therefore, the existence of a one-to-one relationship
in prior observations gives no meaningful indication. In fact, the
qualia themselves may be actively involved in forcing this one-to-one
relationship. Therefore, the presence of a one-to-one relationship
might on the contrary be regarded as evidence for a causal role for
qualia. 

\subsubsection{Counterargument 2: All evidence is tainted by the co-evolution with
qualia}

A second, related counterargument is that that we have no truly admissible
evidence for the one-to-one relationship. All evidence for the one-to-one
relationship comes from real-life circumstances where a causal role
for qualia may have been occurring. Therefore, all of the evidence
has potentially been tainted by the presence of qualia. Furthermore,
even if qualia does have a causal role, looking back at the prior
time-evolution, a one-to-one relationship will be observed which makes
it appear that the qualia are epiphenominal. Therefore, the problem
cannot be decided based upon this relationship in general. 

Because all observations of the past show this one-to-one relationship
regardless of the presence or absence of a causal influence for qualia,
there is no evidence in the natural world available to prove definitively
whether epiphenomenalism is true or false. Therefore, epiphenomenalism
is not provable in principle; it can only be disproven by examining
the statistical likelihoods, and coming to the realization that it
is highly unlikely for the states which occur to arise naturally in
the physical world without a causal mechanism.

Taking a high-level view on the issue, it appears to be the case that
epiphenomenalism has the structure of an unscientific theory. While
it cannot be proven false (because the past time-evolution always
shows the one-to-one relationship), there is no possible evidence
for it (because this one-to-one relationship in principle tells us
nothing about the causal role of qualia), and there are many additional
clues (such as the transmission of information about qualia into physical
world) which make it extremely implausible. 

\subsubsection{Counterargument 3: Novel information}

The case for an epiphenominal one-to-one relationship rests on the
idea that the information present in qualia does not add anything
additional to the information already present in the physical brain
state. We have suggested in previous sections that the information
in the qualia is over and above that present in the brain state based
on three separate grounds. 
\begin{enumerate}
\item Firstly, the qualia may have access to an information processing capacity
beyond the capabilities of the physical brain state (e.g. solving
the binding problem). 
\item Secondly, the information in qualia is of a novel character not predictable
from the underlying physical brain-state (e.g. provides a rich space
of experience, which is i) unique and ii) automatic, and not learned
information). 
\item Thirdly, the information may be of a relational, global type; and
is related to the maintenance of entropic gradients, which support
the persistence and replication of organic life.
\end{enumerate}
Given these considerations, it is established that the information
is present in some form. The remaining point of contention is whether
this information is present in the material world, non-material world,
or perhaps both. 
\begin{enumerate}
\item If the information is an element of the material world, then in the
epiphenominal paradigm it needs to arise out of the material world
through a causal mechanism; for instance it needs to be learned through
the operation of the brain as a physical neural network.
\begin{enumerate}
\item However, it does not appear that this information ever is part of
the material world in the initial state. Therefore it is not clear
where the information arises from. For instance, this information
is not part of the prehistoric conditions of the Earth, prior to the
development of conscious entities, so it is not part of the physical
environment. 
\begin{enumerate}
\item In actuality, the information is hidden in the fine-tuned microcausal
degrees of freedom in the dynamical picture of epiphenomenalism; this
is where the information is arising from in the interpretation. 
\end{enumerate}
\item If the information is not present in the initial state, it is not
clear what is being learned by the biological neural network. The
only thing which can be learned by any learning network is information
encountered in the environment. 
\begin{enumerate}
\item Again, we find that the information spontaneously arises out of the
microcausal degrees of freedom, which implies fine-tuning. 
\end{enumerate}
\end{enumerate}
\item If alternatively the information is just an interpretation or representation
of the material world, and not actually present in the material world,
then possibly there is an escape from the first situation of the information
spontaneously arising. 
\begin{enumerate}
\item However, it is not clear what this information then represents, if
not real information that is present in the material world. 
\begin{enumerate}
\item By analogy we know that words written on a page carry information,
but this information can only be accessed when somebody reads the
words. It doesn't discount the fact that the information is actually
contained in the physical arrangement of words, nor the fact that
the words can play a causal role e.g. the meaning of the words can
change behaviour.
\item This information can a) be stored in non-conscious physical matter
b) can be copied c) be transmitted between individuals by passing
on the physical matter. This would suggest the information is contained
in the physical matter.
\end{enumerate}
\item If we say that the information is not present in the physical matter,
but is in the mind of the observer, this implies that the specific
form and arrangements of matter which corresponds to the information
spontaneously arise. Therefore, it is merely a semantical difference.
We still have no explanation for why these specific forms and arrangements
of matter should spontaneously arise, as they are not predicted by
the conditions of the initial environment. 
\begin{enumerate}
\item The real question is then whether this specific arrangements of matter
in question will cause fine-tuning when the dynamics are reversed;
which is an objective scientific question with a definitive answer.
It can be studied by examining the mathematics and information conservation
properties of the deterministic time-evolution. 
\item We suggest that these specific arrangements of physical matter can
be seen to have information content which is conserved by the time-evolution;
and thus, to conserve this information implies fine-tuning. If fine-tuning
does occur, it can be hypothetically removed, rendering the final
state lacking the specific arrangements of matter in question; thus
proving epiphenomenalism to be false. This argument gets around the
interpretational issues of whether this information is really contained
in the material environment, or instead in the non-material interpretation
or representation of the material environment.
\end{enumerate}
\end{enumerate}
\end{enumerate}

\section{Fine-tuning}

\subsection{What is fine-tuning? }

Fine-tuning can be best understood in the de Broglie-Bohm interpretation.
The de Broglie-Bohm interpretation provides a deterministic dynamics,
based upon the first-order de Broglie guidance equations. The beables
of the de Broglie-Bohm interpretation (e.g. a particle configuration)
have a definite deterministic trajectory in configuration space. If
a final particle configuration is chosen, the motion can be traced
backward in time to find the corresponding initial particle configuration
that will generate this final state. 

The initial particle configuration may appear to be an entirely random
sample from the initial joint probability distribution. However, there
is hidden information contained in the positions of the initial particles,
which is only revealed by operating the deterministic dynamics in
the forward-direction of time once more, whereby this information
becomes manifest and the corresponding final state is revealed. Indeed,
one can engineer the initial particle configuration to give any desired
final state that has a non-zero probability of occurrence, via the
procedure of post-selection of the final state, and then reversing
the dynamics to find the corresponding fine-tuned initial state.

To understand the effect of post-selection further, suppose there
is a game of billiards where the player commenced the game, and by
chance managed to hit every ball into the pockets on a single turn
(an extremely unlikely scenario). If we chose this final state, and
then traced the motion of the billiards backward in time using the
deterministic equations of motion, we would arrive at a set of microcausal
degrees of freedom such as the initial position and velocity of the
cue-ball. 

Just looking naively at the initial position and velocity of the cue-ball,
we might be mislead to believe that there is nothing out of the ordinary.
However, these variables contain a significant amount of hidden information,
as they specify a very unique strike of the ball. When operated in
reverse, the deterministic equations motion allow information of the
final state to be coiled up like a spring into the microcausal degrees
of freedom of the initial state, which are in this case the initial
position and velocity of the cue-ball. This hidden information becomes
manifest over the subsequent forward time-evolution, when it is shown
that the initial strike of the ball results in a very unique final
outcome. This is all a result of the information conservation properties
of the deterministic equations of motion for the system. 

\subsection{Relation to superdeterminism }

Fine-tuning in the de Broglie-Bohm interpretation is superdeterminisic.
The process of fine-tuning causes information from the future state
to be contained in the initial microcausal degrees of freedom, namely
the initial positions of de Broglie-Bohm particles. Under the deterministic
time-evolution, this allows future states to be manufactured, which
consequently violates the free-will assumptions of choices of measurement
basis for example. It is clear that fine-tuning in the de Broglie-Bohm
interpretation can be used to evade Bell's theorem, via the superdeterminism
loophole. However, philosophically this fine-tuning is undesirable,
as future information is contained in the initial microcausal degrees
of freedom. 

\subsection{Philosophical zombies are fine-tuned}

As discussed in previous sections, the procedure of separating the
physical dynamical picture from the non-physical qualia has a counterfactual
post-selection influence on the microcausal degrees of freedom. It
causes them to become fine-tuned, such that they contain hidden information
of the future state, namely that the future state contains a physical
description of functioning brains that appear to have states of consciousness. 

Under a deterministic time-evolution, choosing a final state imposes
an entire trajectory from initial state to final state, due to the
unique mapping between initial and final states. Therefore, one needs
to be careful not to impose fine-tuning on the initial state by making
a selection of the final state. 

Selecting a final state which has the presence of human beings with
physical brains, already is a significant post-selection on the initial
state. It is difficult to avoid post-selection and still have anything
meaningful to discuss about consciousness, because everything we know
about consciousness thus far is grounded in the post-selection of
the final state to resemble the current world where conscious brains
exist. 

\subsection{Removing fine-tuning}

To invalidate the philosophical zombie argument, take the final conditions
e.g. the final particle distribution. Reverse the deterministic dynamics
back to the initial conditions. Now randomly modify the initial microcausal
degrees of freedom, which will ruin the fine-tuning but still be in
accordance with the statistical Born-rule probabilities. Replay the
motion forward in time, and you will inevitably see that the magic
is gone; the philosophical zombie will fail to reproduce the same
statements about their qualia. 

This argument is interesting as it turns determinism upon itself to
disprove the philosophical zombie argument. Clearly the information
about qualia is present at the final state, and it had to have come
from somewhere. However, instead of arguing where it came from, if
we reverse the deterministic time-evolution, the information has to
go back somewhere. The deterministic time-evolution conserves information,
so it cannot just disappear. Consequently, if the information cannot
go back into the qualia (which are now removed from the picture),
the information flows back into the de Broglie-Bohm particle configuration.
Therefore, the particle configuration becomes fine tuned. After removing
the fine-tuning, the information is destroyed permanently, and therefore
it cannot subsequently arise in the final state if the dynamical picture
is played forward in time once more. 

In the first line of argument, we claim that information in the final
state implies fine-tuning in the initial state. In this inverted line
of argument, we claim that given the initial state is fine-tuned,
the fine-tuning can be erased, which also erases the information from
the final state. This is actually a more robust form of argument,
because in the first line of argument, one reaches question of whether
the information really is contained in the physical states of matter
or if it is contained in the mind of the observer. In the second line
of argument, it is clear that something has been erased. The world
cannot be epiphenominal and have this erasure of information. 

This can be understood as a purely mathematical, non-subjective question.
The final arrangement of matter has a mathematically well-defined
information content, based upon a suitable mathematical definition
of information that is conserved. The specific arrangement of matter
in the final state means the information content is higher than otherwise.
Due to information conservation of the time-evolution, the initial
degrees of freedom must be fine-tuned to a greater degree than otherwise.
Removing the fine-tuning decreases the information content of the
initial state, which consequently decreases the information content
of the final state. Clearly something is altered by removing this
information, and the epiphenomenalist account cannot explain how the
information content is decreased by this procedure, or what is removed. 

\subsection{Fine-tuning as a test for consciousness }

Are classical computers capable of consciousness? The Turing test
is not sufficient to answer this question. Classical computers are
a prime example of the philosophical zombie problem. A classical computer
can conceivably tell you it has consciousness, yet on the inside it
is just composed of non-conscious circuits. However, by adapting the
fine-tuning argument above, you will see there evidently may be test
for causally efficacious forms of consciousness, which sheds light
on the problem.

\emph{Steps: }Describe the final state of the computer in detail.
Simulate the dynamics in the reverse direction of time to give the
initial conditions. Perturb the initial conditions slightly, to remove
any fine-tuning. Run the simulation in the forward direction. If the
output of the simulation now lacks any information about experiences
of qualia or consciousness, the original statements about consciousness
likely were genuine, because there was fine-tuning involved, that
has subsequently been removed. Fine-tuning implies hidden information
from qualia was injected into the deterministic process through the
top-down causal channel. However, if the simulation makes the same
statements about consciousness as it did previously, it demonstrates
that fine-tuning was not involved, and the computer was likely not
conscious. 

\subsubsection{Deterministic classical circuits }

A simple application of this principle indicates that classical circuits
made from physical logic gates cannot support causally efficacious
consciousness, because they are robust in the initial conditions and
not capable of fine-tuning. 

Consider a computational circuit made from many dominoes arranged
in a pattern of logic gates. It is evident that the domino computer
is robust in the initial conditions. For instance, slight modification
of the positions of the initial dominoes, e.g. rotating their angles
slightly, will result in the same fall pattern, giving the same computational
output. You could take the final output and trace the initial conditions
back, slightly adjust the positioning of the initial dominoes, run
the device again, and you will get the same answer. Due to this robustness
in the microcausal degrees of freedom, the system is not capable of
becoming fine-tuned to the level at which the fine-tuning affects
the output state. Therefore, there is no room in the physical description
for a causal role for qualia. 

As seen in the above description, fine-tuning appears to be closely
connected to the robustness and information channel capacity of the
input states. What do we mean by robustness? We mean that the initial
dominoes have some redundancy in their physical positioning. You can
move them backwards or forwards slightly or rotate them, and it won't
affect the outcome of the circuit. What do we mean by channel capacity?
The information is a single binary unit traveling down the chain of
dominoes. There is only one real piece of information in each ``wire''
of the circuit, and the microcausal degrees of freedom are irrelevant
to the outcome. 

Due to the limited channel capacity, there are no relevant microcausal
degrees of freedom. The entire system is constructed to be a direct
mapping from inputs to outputs. In particular, there is no cascading
step, where the microcausal conditions get amplified into macroscopic
conditions. Taking all of this into consideration, it implies that
there is no possibility for relevant fine-tuning to be hidden in the
initial arrangements dominoes in the circuit.

\subsubsection{Turing machines are not conscious}

Similar to the domino circuit, it is clear that a hypothetical Turing
machine i.e. an infinite tape with a reading device, cannot have causally
efficacious consciousness. The Turing machine describes a deterministic
set of instructions that is read in sequence. Because the scenario
is abstracted away from the microcausal physical picture as a logical
thought experiment, there are no microcausal degrees of freedom present
at all. Therefore, the system is not capable of fine-tuning. 

One could choose a final state for the Turing machine, then hypothetically
reverse the time-evolution, and reach the corresponding initial state.
Then alter nothing, because there are no microcausal degrees of freedom
available to alter. Run the device again, and exactly the same answer
will be produced. There is only one unalterable path from the initial
conditions to the final conditions. Nowhere in this process can further
causal influences from qualia can be injected into the deterministic
process, because additional causal influences require the possibility
for fine-tuning. 

\subsubsection{How the human brain is able to be conscious}

In the human brain, the complex network of neurons has a cascade of
firings which is highly sensitive to the initial conditions. Essentially,
there needs to be an information bottleneck where the outcome could
be decided one of many ways, and the answer comes down to the microcausal
degrees of freedom. This dependence upon the microcausal degrees of
freedom e.g. the specific arrangement of de Broglie-Bohm particles,
creates the possibility for fine-tuning in the deterministic paradigm,
which allows room for causally efficacious consciousness in the dynamical
picture. 

\subsubsection{How to make a conscious computer}

To generate consciousness within a computer, a necessary condition
is to have a cascading mechanism which will allow for amplification
of fine-tuned microcausal degrees of freedom into macroscopic outcomes.
The computer circuit cannot be a simple deterministic path from macroscopic
inputs to macroscopic outputs. It may be possible to create consciousness
within a computer circuit with the right setup, but current setups
are insufficient as they are based upon classical algorithms with
no indeterminacy or amplification of the microcausal degrees of freedom. 

Standard classical computers cannot not be conscious, because deterministic
classical algorithms are a direct path from inputs to outputs, leaving
no room for additional causal influences to be injected in the forward
picture of the motion, or fine-tuning to occur in the backward picture
of the motion. This rules out the possibility for causally efficacious
forms of consciousness in these systems. 

While we cannot in principle rule out the possibility that deterministic
classical circuits supports an epiphenomenal consciousness, it does
seem highly implausible as it is in disagreement with commonsense
reductionist intuitions that a system which can be decomposed into
deterministic macroscopic parts cannot be conscious. 

However, the requirement for amplification of the microcausal degrees
of freedom is only a necessary condition that can be used to disprove
certain systems are conscious. It cannot be used to prove a system
is conscious, and it is not clear what other requirements are necessary
to make consciousness an inevitable conclusion.

\section{The completeness of quantum mechanics}

\subsection{Born-rule probabilities are a qualia }

The experience of Born-rule probabilities may themselves be regarded
as a form of qualia. The only reason the Born rule probabilities are
observed is because we perceive the world through our senses, including
the qualia of sight of experimental measurement results (for example
observing flashes on a screen during the double-slit experiment). 

However, there are other qualia besides the statistical frequency
of probabilistic events. The rich variety of other qualia also require
explanation. In this sense, quantum mechanics may not be a complete
theory. To complete the quantum description, it might be necessary
to include these other qualia as influences upon the wavefunction
collapse procedure or the de Broglie-Bohm particle configuration. 

\subsection{Born-rule probabilities are not compatible with qualia }

There are branches of the wavefunction which are incompatible with
the Born-rule. Similarly, there are branches of the wavefunction which
are incompatible with qualia. The wavefunction merely sets up a space
of possible worlds, and the vast majority of them are unrealistic.
So the essential question is, ``Are all of the possible Born-rule
worlds compatible with qualia?''. We believe the answer is no. The
worlds compatible with qualia are a subset of the Born-rule worlds.
This is clear because: 
\begin{enumerate}
\item The wavefunction has no knowledge of qualia. Nor does the wavefunction
collapse procedure have knowledge of qualia. Therefore the Born-rule
probabilities have no knowledge of qualia. 
\item Therefore, no information about qualia is imparted into the final
states defined by the Born-rule. 
\item However, it is evident that qualia does impart information into the
final state; therefore, not all Born-rule states can be compatible
with qualia. 
\end{enumerate}
A mechanism is required to ensure only the Born-rule worlds compatible
with qualia are selected. We suggest that the problem may be addressed
by introducing a causal role for qualia, broadly defined. 

\subsection{The intrinsic decoherence connection}

We have highlighted previously in this paper the connection between
the causal role of qualia and the fact that quantum mechanics has
difficulty explaining non-linear and self-referential phenomena. The
quantum formalism typically resorts to a classical-level description
to explain these cases (for instance describing the situation at the
level of classical neurons in the brain). However, in light of the
possible causal role for qualia, it is suggestive that this causal
mechanism may be having the effect of making the quantum state behave
more classically. This enables the effect to be explained without
leaving the quantum description. 

Consequently, we suggest that the role of qualia may be to have a
type of intrinsic decoherence effect, or alternatively a type of quantum
Zeno effect; both of which are mechanisms which would make the quantum
state behave more classically. 

\subsection{The connection to quantum subsystems}

We find it particularly intriguing to suggest a connection between
qualia, which is an internal point of view, and non-Markovianity of
quantum subsystems, which is an incompatibility between the time evolution
of the total quantum system and the time evolution of the subsystem. 

The reason why this might be so is that qualia are related to the
metaphysical property of ``What it's like to be a subsystem?'';
for instance, what it's like to be the `subsystem' of a human brain.
However, the time-evolution of a highly entangled quantum subsystem
is non-Markovian, which is a mysterious property that prevents the
subsystem from having a simple ontological representation. 

Therefore it appears that in quantum mechanics, there is a dichotomy
between the internal point of view from the perspective of subsystems,
and the external point of view from the perspective of the total system.
This mirrors the dichotomy between the internal perspective of subjective
consciousness, and the external perspective which views the brain
in terms of the material physical picture in the context of the environment. 

The phenomenon of non-Markovianity also prevents a reductionist description
of the subsystem, as non-Markovian subsystems describe a complex network
of information flow that cannot be reduced to parts \cite{key-4,key-5}.
The fact that the description is non-reductive may be connected with
consciousness, as it is difficult to envisage that a collection of
individual classical components that have a reductive explanation
(e.g. individual classical neurons) are capable of being conscious. 

Certainly, due to the myriad of interactions occurring, the structures
within the brain are highly entangled, despite being decoherent. Therefore,
it is not out of the realm of possibility that non-Markovianity is
occurring within the brain, and that non-Markovianity plays a role
in consciousness. 

Furthermore, if the total quantum state gets pulled in the direction
of the subsystem; for example, the joint probability density becoming
closer to the product of the marginal probability densities of the
subsystem and environment; that will look like an intrinsic decoherence
is occurring. From the previous discussions on non-linear and self-referential
processes, this is the type of effect which is required, which provides
further evidence that quantum subsystems and non-Markovianity may
be involved in consciousness. 

\subsection{A general causal mechanism?}

We have claimed in previous sections that the causal role for qualia
may allow physical subsystems to violate natural entropy gradients,
and minimize their internal entropy at the expense of the environment.
For example, qualia evidently allow human brains to comprehend information
and meaning, and give rise to an understanding of the physical world
in a global, relational sense. However, deterministic linear dynamical
physical processes are local, and blind to relational information.
Therefore, it is a somewhat paradoxical phenomena potentially resolved
by including a causal role for qualia. 

However, it is evident that the problem is not necessarily restricted
to conscious states or qualia. We ask whether non-conscious can also
access qualia-like features present in the natural world, which enable
this type of behaviour on a smaller scale? Perhaps, this type of information
processing is a general causal mechanism in nature. 

Indeed, the existence of a basic causal mechanism independent of consciousness
seems likely to be the case in light of evolution. It is highly unlikely
that consciousness came about suddenly. More likely, early forms of
life began to use a basic causal mechanism responsible for the effect,
and its use became progressively more refined over the course of evolution.
This argument is interesting, because it brings the debate outside
of the question of consciousness and grounds it in terms of information,
thermodynamics and causality. The debate is a question about the possible
causal mechanisms in nature in general.

These questions are reminiscent early discussions of ``What is life?'',
for instance Schrödinger's seminal paper \cite{key-7} on the topic.
The problem has always been to explain how a living dynamical system
can internally minimize entropy at the expense of the external environment.
Some early conceptions for how this phenomena occurs were proposed,
for instance in the discovery of the formation of spontaneous order
in non-linear systems due to dissipative structures; for which Prigogine
won the 1977 Nobel Prize in Chemistry, and has now evolved into a
large field of study. 

However, we should keep an open mind and recognize that the case may
not be closed. The problem remains that the quantum time-evolution
is linear, and a mechanism for non-linearity beyond the wavefunction
collapse procedure is required to explain the phenomenon. It does
not seem evident prima facie that a linear system can minimise entropy
in this manner. Meanwhile, the potential causal role for qualia has
been overlooked, leaving an opportunity to reexamine the question. 

Perhaps these hypothetical qualia-like effects have allowed biological
systems of all varieties to beat the statistical odds, and remove
entropy from their subsystems, thus creating internal order which
supports self-replication and life. All of these biological systems
are acting like a Maxwell's demon, and we suggest that the explanation
might not be found in the standard linear time-evolution of quantum
mechanics, but in the influence of non-material effects upon the wavefunction
collapse process (for instance via a causal influence of qualia on
the de Broglie-Bohm particle configuration). 

This brings the discussion back to the suggestion that the effect
of qualia is to act as a type of intrinsic decoherence, or alternatively
quantum Zeno-like effect on the quantum state. These mechanisms of
intrinsic decoherence or a quantum Zeno-like effect can easily be
conceived as a general mechanism that operates independently of conscious
experience, and therefore may impact the functions of non-sentient
organic life as well as sentient organic life.

\section{Conclusions}

\noindent This paper has made the case that qualia (i.e. conscious
experiences) have a causally efficacious role, and are not merely
an epiphenomenon. The common view is that the deterministic equations
of motion leave no room for additional causal influences from qualia.
However, such a view contains an error in regarding the dynamics as
a closed form of determinism in the forward direction of the time-evolution. 

The wavefunction is a branching process. Therefore, the type of determinism
is an open form, and there is room in the description for a causal
role of qualia to guide the de Broglie-Bohm particle configuration
down a specific wavefunction branch. Alternatively, this can be understood
as an influence upon the wavefunction collapse process in the Copenhagen
interpretation, as statements made within the de Broglie-Bohm interpretation
have their equivalent in the Copenhagen interpretation. 

\medskip{}
\noindent The evidence for a causal role of qualia is stark. This
evidence includes: 
\begin{enumerate}
\item Arbitrariness of qualia if they do not play a causal role; therefore,
allowing qualia to be hypothetically altered in thought experiments. 
\begin{enumerate}
\item Including philosophical zombie thought experiments. 
\item Including thought experiments where a region of the brain is replaced
with a classical computer circuit. 
\end{enumerate}
These thought experiments violate intuition, and it is not clear that
the physical system can act in the same way upon alteration of the
qualia. 
\item The presence of information about qualia which is able to enter the
physical environment through an unknown channel. 
\begin{enumerate}
\item Including ``zombie island'' thought experiments, where spontaneous
emergence of statements about consciousness occur. 
\end{enumerate}
It is not clear how this information is able to enter the physical
environment without it being sourced from the qualia. 
\item The presence of non-linear and self-referential behavior, despite
the quantum mechanical description being linear and unitary. Noting
that quantum mechanics explains these circumstances by invoking wavefunction
collapse, and therefore relies upon a classical description. 
\item Novel information provided by qualia. 
\begin{enumerate}
\item Including novel information processing capabilities in the brain e.g.
solving the binding problem of consciousness. 
\item Including novel features of qualia e.g. the richness of qualia experiences,
imagined qualia that do not have a direct relation to physical states,
geometrical and/or topological mappings of the space of qualia. 
\end{enumerate}
Furthermore noting that this novel information enters the physical
environment though reports of qualia experiences. 
\item Evolutionary arguments, that possible evolutionary evidence for selection
pressures may be observed in the structure and function of the brain,
which may be an area of future research. 
\item Entropy arguments, that qualia enables physical matter to violate
typical entropy gradients, though the access to relational, top-down
causal information which is not part of the reductionist dynamical
picture.
\end{enumerate}
Meanwhile, the evidence for epiphenomenalism is found to be lacking.
This possible evidence includes: 
\begin{enumerate}
\item The argument that there is a one-to-one correlation between physical
states and qualia, such that any causation is only an apparent causation. 
\begin{enumerate}
\item Where this argument is found to be assuming that which is to be proven,
as it is examining the post-selected, counterfactual state which has
co-evolved with the presence of qualia.
\item Where we recognize that it is impossible in general to remove the
counterfactual post-selection, as all current evidence is based upon
the post-selected state of the current environment that includes qualia. 
\item Where we recognize that the novel information provided by qualia is
not part of the physical environment initially, yet it enters the
physical environment during the time evolution and is present in the
final state. 
\end{enumerate}
\item Philosophical zombie thought experiments. 
\begin{enumerate}
\item Where we find that the microcausal degrees of freedom are fine-tuned,
due to counterfactual post-selection of the final state. 
\item Where we find that removing the fine-tuning in the initial state will
destroy the qualia information in the final state. 
\end{enumerate}
\end{enumerate}
After establishing the evidence for and against epiphenomenalism,
we arrive at the conclusion that qualia must have a causal role in
the dynamics. Using this understanding, we further examine the role
of fine-tuning in the dynamical picture, which occurs if the causal
role of qualia is ignored.

Firstly, the philosophical zombie argument for epiphenomenalism is
found to be insufficient, as it is impossible to separate the material
physical description from the non-material qualia without imposing
counterfactual post-selection, which causes fine-tuning of the microcausal
degrees of freedom. Fine-tuning occurs due to the deterministic dynamics,
which results in a one-to-one mapping between initial states and final
states. Thus choosing a final state which describes philosophical
zombies causes fine-tuning of the initial microcausal degrees of freedom
(e.g. positions of de Broglie-Bohm particles). Undesirable hidden
information about the future state is present in the microcausal degrees
of freedom due to fine-tuning. 

Secondly, fine-tuning is found to be intimately linked to the presence
of causally efficacious consciousness. It is evident that classical
computer circuits; which are incapable of having any meaningful fine-tuning;
are not capable of causally efficacious consciousness. Furthermore,
reductionist arguments against such circuits having epiphenominal
consciousness appear quite strong, therefore it is likely these systems
cannot support consciousness in general. Understanding this link between
consciousness and fine-tuning highlights the difference between classical
circuits and conscious biological brains which can support fine-tuning. 

We continue the discussion by highlighting where the quantum formalism
may not be complete. We show that the Born-rule can be considered
a form of qualia, and that the remaining qualia should not be ignored
in the wavefunction collapse procedure. Furthermore, it is argued
that not all wavefunction branches compatible with the Born-rule are
compatible with the continuous, dynamical experience of qualia across
time. The effect of the causal role for qualia is to select only the
Born-rule states which are simultaneously compatible with qualia. 

Furthermore, it is proposed that qualia may be having a type of intrinsic
decoherence effect, or quantum Zeno effect; both of which would make
the quantum mechanical probabilities appear more classical. This would
resolve the paradox of resorting to a classical-level description
to explain non-linear and self-referential behaviour in quantum mechanics.
Additionally it proposed that qualia and consciousness may be connected
to non-Markovianity in quantum subsystems. This is motivated by the
recognition that qualia is the property of `what it's like to be a
subsystem', whilst non-Markovianity causes physical subsystems to
have a non-reducible ontological description, thus avoiding reductionist
arguments against consciousness. 

Finally, we reflect upon the role that qualia plays in maintaining
entropy gradients in conscious systems, and propose that the effect
may be part of a general causal mechanism, not isolated to the case
of consciousness entities. It is suggested this causal mechanism has
enabled non-conscious organic life to maintain low states of internal
entropy and sustain and replicate itself despite the underlying physical
dynamical picture not having access to top-down, relational information
processing mechanisms. 

\section*{Statements \& declarations}

\subsection*{Statement of originality}

All work is original research and is the sole research contribution
of the author. 

\subsection*{Conflicts of interest: }

No conflicts of interest. 

\subsection*{Funding: }

No funding received. 

\subsection*{Copyright notice: }

Copyright{\small{} }{\footnotesize{}© }2024 Adam Brownstein under
the terms of arXiv.org perpetual, non-exclusive license 1.0.

\end{document}